\documentclass[aps,prl,twocolumn,groupedaddress]{revtex4}
\usepackage{graphicx}
\usepackage{amsmath}
\begin{document}

\title{Normalized General Relativity: Non-closed Universe and Zero Cosmological Constant}
\author{Aharon Davidson}
\email[Email: ]{davidson@bgu.ac.il}
\author{Shimon Rubin}
\email[Email: ]{rubin.shim@gmail.com}
\affiliation{Physics Department, Ben-Gurion University of the Negev, Beer-Sheva 84105, Israel}
\date{March 17, 2013}

\begin{abstract}
We discuss the cosmological constant problem, at the minisuperspace level,
within the framework of the so-called normalized general relativity (NGR).
We prove that the Universe cannot be closed, and reassure that the accompanying
cosmological constant $\Lambda$ generically vanishes, at least classically.
The theory does allow, however, for a special class of $\Lambda \not=0$
solutions which are associated with static closed Einstein universe and
with Eddington-Lema\^{\i}tre universe.

\bigskip
\hspace{-11pt}
\end{abstract}

\maketitle

\begin{center}
\textbf{Introduction}
\end{center}
 \bigskip

Many ideas have emerged over the years in an attempt to decode
the cosmological constant puzzle \cite{Zel'dovich}, clearly one of the most
stubborn problems in contemporary theoretical physics \cite{LambdaRev}.
Among the prominent proposals to explain the cosmological observations
\cite{Riess} one can list unimodular gravity \cite{LambdaSolUni}, alternative
measure gravity \cite{LambdaSolMeasure},  degravitation \cite{LambdaSolDegra},
supersymmetric string theories  \cite{LambdaSolSuper}, and many more.
Most of these proposals, with the exception of unimodular gravity, which
relies on a shift symmetry of the Lagrangian density, differ markedly from
our approach.
The unimodular gravity driven cosmological constant $\Lambda$
turns out to be an arbitrary constant of integration, and one may wonder
whether its value can be fixed, at least classically.
As demonstrated in this paper, this can be done within the framework of
normalized general relativity (NGR) \cite{Tseytlin,NGR}.

NGR is defined by the apparently non-local action
\begin{equation}
         I_{NGR}=\frac{I_{GR}}{\epsilon ~I_{V}}=\frac{\int (-{\mathcal{R}}+\mathcal{L}
        _{m})\sqrt{-g}~d^{4}x}{\epsilon \int \sqrt{-g}~d^{4}x}~\equiv \frac{1\,}{
        \epsilon }\left\langle -\mathcal{R}+\mathcal{L}_{m}\right\rangle \text{,}
\label{Action}
\end{equation}
where the standard Einstein-Hilbert functional $I_{GR}$ is divided by
the volume functional $I_{V}$ such that the underlying theory has a built-in
symmetry 
\begin{equation}
           -\mathcal{R}+\mathcal{L}_{m}\rightarrow {-\mathcal{R}+\mathcal{L}_{m}}+const 
\label{symmetry}
\end{equation}
(with units $16\pi G_{N}=c=\hbar=1$, and the constant $\epsilon$ is introduced
on purely dimensional grounds).
The field equations associated with the action Eq.(\ref{Action}) take the form 
\begin{equation}
          \mathcal{R}_{\mu \nu }-\frac{1}{2}g_{\mu \nu }\mathcal{R}=\Lambda g_{\mu \nu
          }-\frac{1}{2}T_{\mu \nu },  
\label{NGReq}
\end{equation}
where the cosmological constant $\Lambda$
is proportional to the value of the action along the classical solution. Furthermore, 
$\Lambda$ is determined by the self-consistency functional equation
\begin{equation}
           \Lambda=\frac{\epsilon}{2 } I_{NGR}(\Lambda ),
\label{Ieq}
\end{equation}
where $I_{NGR}(\Lambda )$ stands here for the on shell value of $I_{NGR}$,
i.e., its value along the accompanying 
classical equations of motion Eq.(\ref{NGReq}).
Together, Eq.(\ref{NGReq}) and Eq.(\ref{Ieq}) are non-local (in the sense
defined in \cite{Gel'fand}) and acasual, since they contain a spacetime
average of the Einstein-Hilbert Lagrangian.
Roughly speaking, NGR cannot be solved as an ordinary Cauchy problem.
Yet once the functional equation 
Eq.(\ref{Ieq}) is solved and the corresponding $\Lambda$ gets fixed,
Eq.(\ref{NGReq}) becomes practically local.
Resembling NRF, the practically local equations of motion which stem from
our NGR Lagrangian do not allow for any flow of information from the future
whatsoever. Invoking the global structure of the manifold in order to
self-consistently fix the value of an integration constant needs not be
generically interpreted as non physical.
On the contrary, respecting the NRF mathematical consistency, the average
of the Ricci curvature must be carried out over the entire space-time manifold,
and thus should be viewed as nothing but a mathematical self consistency
requirement.

Notice that, as expected on symmetry grounds, where the action $\tilde I_{NGR}$
differs from the action $I_{NGR}$ by just one constant
\begin{equation}
          \tilde{I}_{NGR}\equiv \frac{1\,}{\epsilon }\left\langle -\left( \mathcal{R}+2\Lambda
           _{0}\right) +\mathcal{L}_{m}\right\rangle=I_{NGR}-\frac{2\Lambda_{0}}{\epsilon},
\label{NGRwithConst}
\end{equation}
we expect that classically, $\tilde{I}_{NGR}$ and $I_{NGR}$ lead to the same solution.
Indeed, in case we treat $\tilde{I}_{NGR}$
as our basic action,  a seemingly different equations of motion emerges, given by
\begin{equation}
          \mathcal{R}_{\mu \nu }-\frac{1}{2}g_{\mu \nu }\mathcal{R}=\Lambda_{p} g_{\mu \nu
          }-\frac{1}{2}T_{\mu \nu },
\label{NGReqLambda0}
\end{equation}
where $\Lambda_{p}$ is defined as $\Lambda+\Lambda_{0}$.
The latter expression serves here as the newly emergent effective combination for the cosmological constant,
and it is built from two contributions: $\Lambda_{0}$ due to the constant we
did not drop in the action Eq.(\ref{NGRwithConst}), and $\Lambda$ is proportional to $\tilde I_{NGR}$.
Nevertheless, we can easily see that Eq.(\ref{NGRwithConst}) and Eq.(\ref{NGReqLambda0}) imply that the constant $\Lambda_{p}$ 
is subject to the functional equation
\begin{equation}
          \Lambda_{p}=\frac{\epsilon}{2 } I_{NGR}(\Lambda_{p}),
\end{equation}
which is identical to Eq.(\ref{Ieq}).
To summarize, adding a constant $\Lambda_{0}$ has no effect
whatsoever on the physical solution, and in practice, the solution tends toward the same value of the physical cosmological constant. 
Consequently, assigning different values to the constant $\Lambda _{0}$ 
in the numerator of Eq.(\ref{Action}) does not generate spacetime curvature. In fact, empty spacetime 
($\mathcal{L}_{m}=0$) in NGR necessarily leads to the Ricci flat Universe, $\mathcal{R}_{\mu \nu }=0$, and to a
vanishing cosmological constant. It differs from
the theory of general relativity, where starting with an empty spacetime and
various values of $\Lambda _{0}$ in the Einstein-Hilbert action, produces correspondingly
differently curved spacetimes. Einstein's discontent with this feature of his own
theory, represented by the de Sitter solution, can be found in \cite{Pais}.

For a number of other relatively simple cases, such as a point
particle and a minimally coupled quintessential scalar field, NGR
leads to a vanishing cosmological constant \cite{Tseytlin}, \cite{NGR} and 
\begin{equation}
          \left\langle \mathcal{R}\right\rangle =0.  
\label{Flat}
\end{equation}
Nevertheless, deviating from the normalized Einstein-Hilbert Lagrangian can
drastically change Eq.(\ref{Ieq}), and give rise to a non-trivial $\Lambda $
and $\left\langle \mathcal{R}\right\rangle \not=0$, even in the absence of
matter fields \cite{NGR}. For instance, starting from ``Normalized $F(\mathcal{R})$ gravity", 
described by the action $\left\langle F(\mathcal{R})\right\rangle $,
one can show that Eq.(\ref{Ieq}) written for a constant curvature
manifold, $\mathcal{R=}const$, acquires the non-trivial form 
\begin{equation}
           \left\langle \mathcal{R}\frac{dF}{d\mathcal{R}}\right\rangle =0\text{.}
\end{equation}
Considering such deviation, convinces that the normalized
Einstein-Hilbert action is unique, in the sense that the action for
which $\left\langle \mathcal{R} \right\rangle=0$ is the only solution that 
follows for the case in which the spacetime is empty. This suggests that the consideration of
more general normalized action principles (to be discussed elsewhere) may be related to the secondary cosmological 
constant puzzle \cite{CCvalue}, namely, "Why is there something rather than nothing?" 
Keeping in mind that determining $\Lambda$, i.e, solving Eq.(\ref{Ieq}), for the case of 
an arbitrary matter Lagrangian and underlying symmetry, is difficult
to address directly, here we focus on a maximally symmetric case.

The rest of this work is organized as follows: first, we review some general
features of NGR. Afterwards, we formulate its
minisuperspace version, written for the scale factor function, 
the only degree of freedom in the corresponding Friedmann-Robertson-Walker (FRW) line element.
The subsequent investigation of the associated equations of motion and the accompanying functional equation
leads us to the conclusion that the case of the spatially closed Univese does not respect 
the corresponding functional equation Eq.(\ref{Ieq}) for conventional attractive matter. 
Therefore, adopting the NGR paradigm 
in a maximally symmetric FRW Universe with attractive matter leads toward a non-closed Universe.
Furthermore, we reconfirm this conclusion in the framework of a somewhat different 
action principle \cite{Jackiw} designed to take into account the dynamics of the relativistic 
perfect fluid matter.

\bigskip

\begin{center}
\textbf{Minisuperspace Model in Normalized General Relativity}
\end{center}

 \bigskip

Implementing the cosmological principle of spatial homogeneity and isotropy
for the Universe at the large scale \cite{Wald} leads to the description of the spacetime metric by just
one scale factor function. In this maximally symmetric case, the most general line element 
in co-moving coordinates is the FRW line element explicitly given by 
\begin{equation}
           ds_{FRW}^{2}=-dt^{2}+a^{2}(t)\left( \frac{dr^{2}}{1-kr^{2}}+r^{2}d\Omega^{2}\right),  
\label{T}
\end{equation}
where $k=-1,0,1$, corresponding to a spatially open, flat, or closed universe,
respectively. The cosmic time evolution of the scale factor $a(t)$ is
governed by the Einstein equation 
\begin{equation}
           \frac{\dot{a}^{2}+k}{a^{2}}=\frac{1}{3}\left(\frac{1}{2} \rho +\Lambda \right),
\label{Motion}
\end{equation}
where $\rho (a)$ denotes the energy density of a perfect fluid
energy-momentum tensor 
\begin{equation}
           T_{\nu }^{\mu }=diag\left( -\rho ,p,p,p\right),  
\label{Te}
\end{equation}
and $\Lambda $ is the still undetermined cosmological constant. The
various components of the energy-momentum tensor are subject to some yet
unspecified equation of state $p=p(\rho )$, and must obey the
local conservation law 
\begin{equation}
          \dot{\rho}+3\frac{\dot{a}}{a}(\rho +p)=0.  
           \label{Conservation}
\end{equation}

Invoking the minisuperspace formalism to study normalized general relativistic cosmology, our
starting point is the action 
\begin{equation}
             I=\frac{1}{\epsilon }\left\langle 6\frac{\ddot{a}a+\dot{a}^{2}+k}{a^{2}}
            -\rho (a)\right\rangle.  
\label{ActionNGR}
\end{equation}
While the kinetic component is easily recognized as the Ricci scalar curvature
term $\mathcal{R}$, the tenable potential part of the energy density, $\rho (a)$,
is actually introduced by hand. 
The integrations over $3$-space coordinates in the numerator and denominator of action Eq.(\ref{ActionNGR})
share the same domain and are eventually reduced.
The vanishing variation of the action $I$ with respect to the scale factor $a$
\begin{equation}
           \delta \int \left( 6\frac{\ddot{a}a+\dot{a}^{2}+k}{a^{2}}-\rho \left(
            a\right) \right) {a^{3}}dt-2\Lambda \delta \int a^{3}dt=0  \label{Variation}
\end{equation}
gives rise to the following equations of motion
\begin{equation}
           2\ddot{a}a+\dot{a}^{2}+k=\left( \Lambda +\frac{1}{2}\rho \left( a\right) \right) a^{2}+ 
           \frac{a^{3}\left( a\right) }{6}\frac{d\rho}{da}
\label{MotionSecond}
\end{equation}
where according to Eq.(\ref{Ieq}), $\Lambda $ has been identified via 
the functional equation
\begin{equation}
            \Lambda=\frac{\epsilon }{2}I(\Lambda).
\label{FuncEq}
\end{equation}
The key point is that to construct an explicit expression for the functional equation Eq.(\ref{FuncEq}), 
the associated classical solution of the equations of motion, $a\left(t;\Lambda \right)$, must be plugged back into Eq.(\ref{ActionNGR}). 
In fact, utilizing the equations of motion Eq.(\ref{Motion}) and
Eq.(\ref{MotionSecond}) to eliminate $\dot{a}$ and $\ddot{a}$ in the action Eq.(\ref{ActionNGR}) gives
\begin{equation}
          \Lambda =-\frac{1}{2}\left\langle \rho \left( a\right) +\frac{1}{2}\frac{d\rho(a)}{da} a\right\rangle.
\label{LambdaRhoRhoPrime}
\end{equation}
Here the cosmological constant $\Lambda$ enters the equation through the scale factor and may also appear explicitly
in the time integration limits.
At this point we can rewrite the functional equation Eq.(\ref{LambdaRhoRhoPrime}) in one of the two following ways:
(i) utilizing the equations of motion Eq.(\ref{Motion}) and Eq.(\ref{MotionSecond}) to eliminate $\rho$ and $d\rho/da$ and obtain
\begin{equation}
           \left\langle \frac{\ddot{a}}{a}\right\rangle=0,
\label{Kinematic}
\end{equation}
or (ii) utilizing the conservation law Eq.(\ref{Conservation}) to eliminate $d\rho/da$  in Eq.(\ref{LambdaRhoRhoPrime}), which yields
\begin{equation}
           \Lambda =\frac{1}{4}\left\langle \rho (a)+3p(a)\right\rangle
\label{LambdaRhoP}.
\end{equation}
Notably, the combination $\rho +3p$ is a
source of geodesic acceleration in a Raychaudhuri equations \cite{Wald} that measures the relative acceleration
of two geodesics in the spacetime. Its spatial component $\vec{g}$, which is subject to $\vec{\nabla}\cdot \vec{g}=-\frac{1}{4}\left( \rho +3p\right)$, merely implies that gravity is attractive if $\rho +3p>0$. By virtue of Eq.(\ref{LambdaRhoP}), 
this fact alone implies that the corresponding solution of the functional equation Eq.(\ref{LambdaRhoP}), if it exists,
must be non-negative.

One technical remark is now in order: the first equation of motion Eq.(\ref{Motion})
may be derived from Eq.(\ref{MotionSecond}) by multiplying the latter by $\dot{a}$
and then time integrating the result. This gives rise to Eq.(\ref{Motion}) but with an additional superfluous integration constant,
a fictive indeterminacy that can be avoided once we fix the lapse function $N(t)$ introduced via
\begin{equation}
           ds^{2}=-N(t)^{2}dt^{2}+a^{2}(t)\left( \frac{dr^{2}}{1-kr^{2}}+r^{2}d\Omega
           ^{2}\right),
\label{Lapse}
\end{equation}
after the variation of the corresponding minisuperspace normalized action
\begin{equation}
             I=\frac{1}{\epsilon }\left\langle 6\frac{N(\ddot{a}a+\dot{a}^{2})+kN^{3}-a\dot{a}\dot{N}}{a^{2}N^{3}}
            -\rho (a)\right\rangle.  
\label{ActionNGRlapse}
\end{equation}
In so doing, the variation with respect to $N(t)$ and $a(t)$,
combined with the gauge choice $N(t)=1$ implemented after the variation,
leads to Eq.(\ref{Motion}) and Eq.(\ref{MotionSecond}), respectively.

Further determination of $\Lambda $ requires the use of a perfect fluid equation of state.
The equation
\begin{equation}
          p=w\rho.
\label{PFEqState}
\end{equation}
is often a sensible starting point in standard cosmology \cite{Wald}. 
Here $w$ is a constant usually bounded between unity and minus unity.
The corresponding energy density $\rho$ is easily found from the underlying conservation law, Eq.(\ref{Conservation}), 
\begin{equation}
           \rho\left( a\right) = 6c^{2}a^{-3\left(1+w\right)}, 
\label{rhoPF}
\end{equation}
where $c$ is a constant and the factor of six was introduced for convenience.
Substituting the latter into Eq.(\ref{LambdaRhoP}) allows the functional equation for $\Lambda$
to be rewritten
\begin{equation}
           \Lambda =\frac{3c^{2}(1+3w)}{2}\left\langle a\left( t;\Lambda \right)
           ^{-3\left( 1+w\right) }\right\rangle
\label{GeneralEquationState}
\end{equation}
in terms of the scale factor alone.
The sign of $\rho +3p$ dictates that we check the self-consistency of the following example cases:

\bigskip

  (i) \hspace{0.3pt} $\rho +3p>0$, \hspace{2pt} e.g., \hspace{2pt} $\rho=\frac{6c^{2}}{a^{4}}$, \hspace{2pt} $p=\frac{1}{3} \rho \Longrightarrow \Lambda \geq 0$

 (ii) \hspace{0.1pt} $\rho +3p=0$, \hspace{2pt} e.g.,\hspace{2.5pt} $\rho=\frac{6c^{2}}{a^{2}}$, \hspace{2pt} $p=-\frac{1}{3} \rho \Rightarrow \Lambda = 0$

(iii) $\rho +3p<0$, \hspace{2pt} e.g., \hspace{0.1pt} $\rho=\frac{6c^{2}}{a}$, \hspace{1pt} $p=-\frac{2}{3} \rho \Rightarrow \Lambda \leq 0$.

\bigskip 

\hspace{-13pt} As we will see below, the functional equation Eq.(\ref{GeneralEquationState}) does not possess a solution for any choice of $k$ or $w$. 

\bigskip

\noindent \textbf{(i)} \textit{\textbf{Conventional matter}} $\rho +3p>0$.

\bigskip

We now consider the solution of Eq.(\ref{GeneralEquationState}) for $\Lambda $ in the
specific case of a radiation dominated Universe, characterized by $w=1/3$ and $\rho\left(a\right) = 6c^{2}/a^{4}$.
Choosing other possible values of $w$ in the range $w>-1/3$, which
is consistent with the underlying strong energy condition, does not alter our conclusions markedly.
In general, as we will confirm below, the expression given by Eq.(\ref{GeneralEquationState}) vanishes in the limit 
the scale factor $a$ grows unbounded. 
Consequently, by using Eq.(\ref{GeneralEquationState}), the corresponding solution for the cosmological constant
must also vanish. The strictly positive solution, on the other hand, may occur only if the spacetime volume is finite. 

To find the solution of the functional equation, Eq.(\ref{GeneralEquationState}), we can solve for the zeros the difference function, $D(\Lambda)$, defined as a difference between the two sides
of Eq.(\ref{GeneralEquationState})
\begin{equation}
         D(\Lambda)=\frac{3c^{2}(1+3w)}{2}\protect\int \limits_{t_{-}}^{t_{+}} a^{-3w}dt/\protect \int\limits_{t_{-}}^{t_{+}}a^{3}dt-\Lambda.
\label{DLambda}
\end{equation}
Here the scale factor $a$ and the time limits $t_{+}$, $t_{-}$ depend on $\Lambda$. 
Keeping in mind that Eq.(\ref{LambdaRhoP}) and Eq.(\ref{GeneralEquationState}) imply 
that the associated $\Lambda$ in the present case is positive or zero, we address these two cases separately below.

\bigskip

(i1) \textit{Is $\Lambda>0$ possible?}

\bigskip

For this case, the corresponding equation of motion, Eq.(\ref{Motion}), takes the following
form of a non-relativistic mechanical problem 
\begin{equation}
          \frac{1}{4}\dot{x}^{2}+\left( kx-\frac{\Lambda }{3}x^{2}\right) =c^{2}.
\label{Potential}
\end{equation}
For convenience, we introduced a new variable $x=a^{2}$
and consider $U(x;\Lambda)$ defined by
\begin{equation}
          U(x;\Lambda)=kx-\frac{\Lambda}{3}x^{2},
\label{MechPoten}
\end{equation}
as corresponding to the one parameter family of mechanical potentials with $\Lambda$ serving as a parameter.
This results in useful expressions for the classical turning points $x_{\pm }$
\begin{equation}
           x_{\pm }=\frac{3}{2\Lambda }\left( k\pm \sqrt{k^{2}-\frac{4\Lambda c^{2}}{3}}\right),
\label{turning}
\end{equation}
and leads the present discussion to address:

\bigskip

(i1.1) $k\leq 0$

(i1.2) $k>0$ and $\displaystyle{k^{2}<\frac{4\Lambda c^{2}}{3}}$

(i1.3) $k>0$ and $\displaystyle{k^{2}\geq \frac{4\Lambda c^{2}}{3}}$.

\bigskip

(i1.1) In this case the classically allowed region for $x$ contains all
non-negative values, and consequently, the spacetime volume grows unbounded.
Because the averaged quantity in Eq.(\ref{GeneralEquationState}) is to a negative power,
we expect the average itself, i.e., a quotient of two integrals, to tend to zero as the scale factor grows.
In fact, this behavior may be observed numerically once we consider the quantity $\left\langle \rho (a)+3p(a)\right\rangle_{T}$, defined by
\begin{equation}
           \left\langle \rho (a)+3p(a)\right\rangle_{T}= \protect\int \limits_{0}^{T}\left( \protect\rho(a) +3p(a)\right) a^{3}dt/\protect
\int \limits_{0}^{T}a^{3}dt.
\end{equation}
Here $\rho(a)$ and $p(a)$ are evaluated along the corresponding classical solution
$a(t;\Lambda)$ with an as yet unspecified $\Lambda \geq 0$ and with the parameter $T$ serving as an upper cutoff for the time coordinate.
In FIG. 1 we show that the resultant $\left\langle \rho (a)+3p(a)\right\rangle_{T}$ vanishes as $T \rightarrow \infty$.
\begin{figure}[th]
          \includegraphics[scale=0.45]{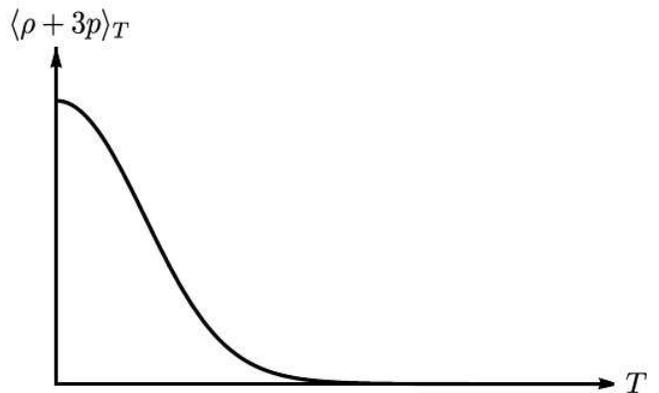}
          \caption{Typical profile of the ratio $\left\langle \protect\rho +3p\right\rangle _{T}$ as a function of $T$ along the
                        classical solution $a(t;\Lambda)$ with $\Lambda > 0$ in the case (i1.1). In this case $\left\langle \protect\rho +3p\right\rangle _{T}$ tends to zero 
                         as $T\rightarrow \infty$.}
\label{fig1}
\end{figure}

To summarize (i1), the zero limit singles out the only possible value of
the cosmological constant, which is $\Lambda=0$.
Since this value is not compatible with the initial $\Lambda>0$ assumption,
we conclude that the expression given by Eq.(\ref{DLambda}) does not vanish;
therefore, this case does not admit a solution.

\bigskip
(i1.2) The spacetime volume in this case is also unbounded; therefore, we would
expect that the accompanying $\Lambda $ is driven to zero. However, this implies that the
relation $k^{2}<4\Lambda c^{2}/3$
is eventually reduced to $k^{2}<0$, which does not hold for any real $k$. Therefore, the present case 
is not allowed since it contradicts the underlying initial assumption.
This point may also be seen graphically in FIG. 2. In fact, the graphs of the corresponding one-parameter family of mechanical potentials given by Eq.(\ref{MechPoten}) lay below the total mechanical energy. 
This implies that the classical region is infinite, which is not compatible with the initial assumption of $\Lambda >0$.

\bigskip
(i1.3) In this case we will find the allowed value of $\Lambda$ by utilizing
Eq.(\ref{Kinematic}), afterward providing an exact calculation based on Eq.(\ref{DLambda}).
 
 As may be seen in FIG. 2, the allowed region for the variable $x$ in this case consists of two disjointed regions $0\leq
x\leq x_{-}$ and $x_{+}\leq x\leq \infty$. The corresponding classical solution in the region $x_{+}\leq x\leq \infty$ 
possesses a diverging spacetime volume, and therefore, the associated value of $\Lambda$ necessarily vanishes.
This, however, is not consistent with the underlying assumption, i.e., $k>0$ and $k^{2}\geq 4\Lambda c^{2}/{3}$, which excludes the possibility of $\Lambda=0$. 

The classical solution in the region $0\leq x\leq x_{-}$, on the other hand, 
has a finite spacetime volume, and in combined with Eq.(\ref{GeneralEquationState}), it implies that possible values of $\Lambda$ are positive.
\begin{figure}[th]
\includegraphics[scale=0.4]{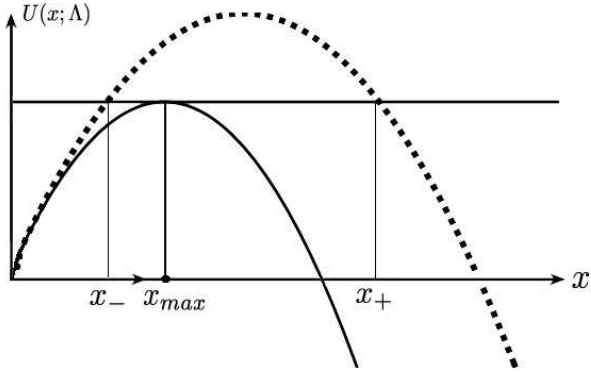}
\caption{The mechanical potential $U(x;\Lambda)$ 
             is plotted for two different non-zero values of $\Lambda$,
             together with the corresponding total mechanical energy $c^{2}$  (horizontal line). 
             The only consistent value of $\Lambda$ with Eq.(\ref{Kinematic}) in the present case, is that leading to the mechanical potential (solid line)
             tangent to the total mechanical energy at point $x_{max}$. The associated solutions correspond to
             the Eddington-Lemai\^{t}re Universe (arrow) and its Einstein static Universe limit (solid point).}
\label{fig2}
\end{figure}
The specific value of $\Lambda$ which solves the functional equation Eq.(\ref{Kinematic})
must result either in a non-constant $a$ such that $\ddot{a}\rightarrow 0$ as $t\rightarrow \infty $ or in $\ddot{a}\equiv0$
associated with an expanding Eddington-Lema\^{\i}tre Universe 
\cite{Eddington} and its Einstein static closed Universe limit, respectively.
The corresponding value of $\Lambda$ dictates $x_{+}=x_{-}$, i.e., the graph of $U(x;\Lambda)$ 
is tangent to the graph of the total mechanical energy  (see FIG.2), and eventually leads to 
\begin{equation}
           \Lambda =\frac{3}{4c^{2}},
\label{caseithree}
\end{equation}
once $k=1$ has been substituted.

To verify that the only allowed value of $\Lambda$ in the present case is given by Eq.(\ref{caseithree}),
we can plot the graph of the function $D(\Lambda)$ given by Eq.(\ref{DLambda}). 
Specifically, plugging the solution
of Eq.(\ref{Potential}) (written with the variable $a$) given by
\begin{equation}
\begin{split}
           a(t;\Lambda)=\sqrt{\frac{3k}{2\Lambda}}
                     &\left(1-\sinh\left( \sqrt{\frac{4\Lambda}{3}t}\right)+ 
\right.
\\
&\left.        
                  \sqrt{\frac{4 \Lambda c^{2}}{3k^{2}}}
                  \cosh\left( \sqrt{\frac{4\Lambda}{3}t}\right)\right)^{\frac{1}{2}},
\end{split}
\end{equation}
into Eq.(\ref{DLambda}) leads to
\begin{equation}
         D(\Lambda)=3c^{2}\frac{\protect\int \limits_{t_{-}}^{t_{+}}\displaystyle{\frac{dt}{a}}}
       {\protect\int \limits_{t_{-}}^{t_{+}}a^{3}dt}-\Lambda,
\label{PosDifference}
\end{equation}
where the integration limits are given by
\begin{equation}
\begin{split}
        &t_{-}=0
\\
        &t_{+}=\frac{1}{4}\sqrt{\frac{3}{\Lambda}}\log\left(\left(1+\frac{2c}{k}\sqrt{\frac{\Lambda}{3}} \right)/\left(1-\frac{2c}{k}\sqrt{\frac{\Lambda}{3}} \right)\right).
\end{split}
\end{equation}
Plotting the corresponding graph of the function $D(\Lambda)$ for $w=1/3$ and
$k=1$ indicates that the root of the function $D(\Lambda)$ in the present case can be found
only for the value given by Eq.(\ref{caseithree}).
\begin{figure}[th]
          \includegraphics[scale=0.25]{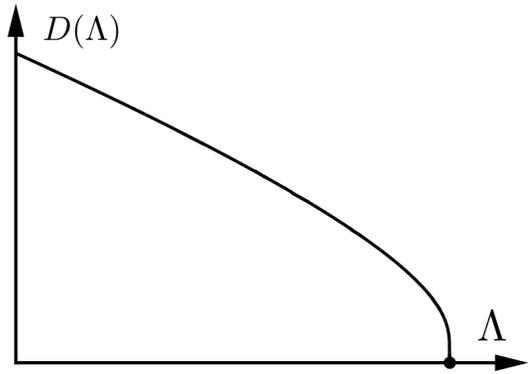}
          \caption{The graph of the function $D(\Lambda)$ from Eq.(\ref{PosDifference}) and its 
sole root given by Eq.(\ref{caseithree}).}
\label{fig3}
\end{figure}
 
Unfortunately, the value of the cosmological constant given by Eq.(\ref{caseithree}) exhibits a singular behavior in
the limit of vanishing matter characterized by the $c\rightarrow 0$ limit. This contradicts one of the  
NGR results, i.e., the cosmological constant necessarily vanishes in an empty spacetime \cite{Tseytlin}, \cite{NGR}.
Therefore, we exclude these cases as possible physical solutions at this stage.

To conclude, the only possible solution in this category is characterized by a positive cosmological constant, given by Eq.(\ref{caseithree}),
and associated with an eternally expanding Eddington-Lemai\^{t}re solution and its static Einstein Universe 
limit.

\bigskip

(i2) \textit{Is $\Lambda=0$ possible?}

\bigskip

To investigate this case, we solve Eq.(\ref{Potential}) with $\Lambda=0$, given by
\begin{equation}
          \frac{1}{4}\dot{x}^{2}+kx=c^{2},
\label{Potential2}
\end{equation}
and then insert the corresponding solution into Eq.(\ref{GeneralEquationState}).
Consequently, the discussion of this case addresses:

\bigskip

(i2.1) $k>0$; \hspace{1pt}  $a(t)=\sqrt{2ct-kt^{2}}$,  \hspace{13pt} $0 \leq t \leq 2c/k$

(i2.2) $k=0$; \hspace{1pt} $a(t)=\sqrt{2ct}$, \hspace{39pt} $0 \leq t \leq \infty$

(i2.3) $k<0$; \hspace{1pt} $a(t)=\sqrt{2ct+\left| k \right|t^{2}}$, \hspace{4pt} $0 \leq t \leq \infty$,

\bigskip

\hspace{-15pt} where we have provided the accompanying solutions of Eq.(\ref{Potential2}), written explicitly with the scale factor $a$.
Plugging each of those solutions into Eq.(\ref{DLambda}) leads to the following conclusions:
\bigskip

(i2.1)  The corresponding value of $D(0)$ is explicitly given by
\begin{equation}
         D(0)=3\frac{\protect\int \limits_{0}^{\frac{c}{k}} \frac{c^{2}}{\sqrt{2ct-kt^{2}}}dt}{\protect\int \limits_{0}^{\frac{c}{k}}\left( \sqrt{2ct-kt^{2}}\right)^{3}dt}=\frac{8 k^{2}}{c^{2}} \neq 0.
\end{equation}
Therefore, we are led to the conclusion that the $k>0$ case is not compatible in a Universe with a vanishing
cosmological constant and attractive matter.
\bigskip

 (i2.2) The corresponding value of $D(0)$ is explicitly given by
\begin{equation}
          D(0)=3\frac{\protect\int \limits_{0}^{T} \frac{c^{2}}{\sqrt{2ct}}dt}{\protect\int \limits_{0}^{T}\left( \sqrt{2ct}\right)^{3} }=\frac{152}{8T^{2}} \rightarrow 0;
\quad \mbox{as} \quad T \rightarrow \infty.
\end{equation}
This leads us to the conclusion that the $\Lambda=0$ case is consistent with the $k=0$ flat Universe with attractive matter. 
\bigskip

(i2.3)  The corresponding value of $D(0)$ is explicitly given by
\begin{equation}
\begin{split}
          D(0)=3&\frac{\protect\int \limits_{0}^{T} \frac{c^{2}}{\sqrt{2ct+\left| k \right|t^{2}}}dt}{\protect\int \limits_{0}^{T}\left( \sqrt{2ct+\left| k \right|t^{2}}\right)^{3}dt} \simeq 
\\
-\frac{12 c^{2}}{k^{2}T^{4}}&\log{\left(4\left| k \right|\sqrt{-T}\right)} \rightarrow 0;
\quad \mbox{as} \quad T \rightarrow \infty.
\end{split}
\end{equation}
This leads us to the conclusion that the $\Lambda=0$ case is consistent with the $k<0$ open Universe with attractive matter.
\bigskip

To summarize the case of attractive matter, characterized by $\rho+3p>0$, 
we conclude that the only possible solutions correspond to $\Lambda=0$ and $k \leq 0$ ; 
$\Lambda=3/4c^{2}$ and $k=1$.
The solutions with a strictly positive cosmological constant, are not physcially viable,
and the generic solution corresponds to the non-closed Universe with a vanishing cosmological constant.

\bigskip

\noindent \textbf{(ii)} \textit{\textbf{String-like network}} $\rho +3p=0$. 

\bigskip

The functional equation, Eq.(\ref{GeneralEquationState}), implies
that the perfect fluid, with a $w=-1/3$ equation of state and $\rho \left( a\right) =6c^{2}/a^{2}$,  leads to a vanishing
cosmological constant. This fact does not depend on the sign of the underlying spacetime
curvature, and one just has to verify whether or not any limitations are implied by the equation of motion, Eq.(\ref{Motion}), itself.
In our case, the latter takes the following form
\begin{equation}
           \dot{a}^{2}=\frac{c^{2}}{3}-k,
\label{Potential3}
\end{equation}
and includes practically no limitations on the sign of the spatial curvature of the Universe.
In fact, from the equation of motion Eq.(\ref{Potential3}), we learn that the cases $k<0$ and $k=0$ are allowed without any restrictions, while the case $k>0$ is allowed if $c^{2}>3$ holds. 

\bigskip

\noindent \textbf{(iii)} \textit{\textbf{Ghosty matter}} $\rho +3p<0$. 

\bigskip

In this case Eq.(\ref{LambdaRhoP}) implies that the allowed values of the cosmological constant
are necessarily non-positive. Taking, for instance, $\rho\left( a\right) =-6c^{2}/a$, 
which corresponds to $w=-2/3$, leads to
the following equation of motion for the scale factor $a$
\begin{equation}
           \dot{a}^{2}-\frac{\Lambda}{3}a^{2}-c^{2}a=-k.  
\label{Potential3_1}
\end{equation}
In the current case we briefly state the final results, which can be obtained by
following the same steps as above.

\bigskip

(iii1) \textit{Is $\Lambda<0$ possible?}

\bigskip

In this case Eq.(\ref{Potential3}) takes the form
\begin{equation}
           \dot{a}^{2}+\frac{\left|\Lambda\right|}{3}a^{2}-c^{2}a=-k
\label{Potential4}
\end{equation}
and when treated as a one-dimensional mechanical problem, it admits the following classical turning points
\begin{equation}
           a_{\pm}=\frac{3c^{2}}{2\left|\Lambda\right|}\left(1\pm\sqrt{1-\frac{3k}{4c^{4}}\left|\Lambda\right|}\right).
\end{equation}
Similar to the cases considered above, the present discussion addresses the following cases:

\bigskip

(iii1.1) $k>0$, $k>3c^{2}/4\left|\Lambda\right|$; not allowed

(iii1.2) $k>0$, $k=3c^{2}/4\left|\Lambda\right|$; allowed with $\Lambda=-3c^{2}/4$

(iii1.3) $k<0$, $k>3c^{2}/4\left|\Lambda\right|$; not allowed

(iii1.4) $k=0$;  \hspace{21 mm} not allowed

(iii1.5) $k<0$;  \hspace{21 mm} not allowed

\bigskip
\hspace{-13pt} The validity of each case may be obtained once we follow practically the same steps as above.
A short analysis reveals that the only possible solution consistent with Eq.(\ref{Kinematic}) 
is the static Universe with $k=-1$ and 
\begin{equation}
          \Lambda =-3c^{2}/4.
\label{NegLam}
\end{equation}
To see this explicitly, we consider the difference, $D(\Lambda)$, given by Eq.(\ref{DLambda})
\begin{equation}
         D(\Lambda)=\left| \Lambda\right|-\frac{3c^{2}}{2}\frac{\protect\int \limits_{a_{-}}^{a_{+}}\frac{a^{2}da}{\sqrt{-\frac{\left| \Lambda \right|}{3}a^{2}+c^{2}-k}}}
       {\protect\int \limits_{a_{-}}^{a_{+}}\frac{a^{3}da}{\sqrt{-\frac{\left| \Lambda \right|}{3}a^{2}+c^{2}-k}}},
\label{Difference}
\end{equation}
and seek the corresponding values of $\Lambda=-\left| \Lambda \right|$ such that $D(\Lambda)$ vanishes.
Performing numeric integration one can plot the resulting graph of $D(\Lambda)$ (see FIG. 4)
where the limits $a_{\pm}$ are given by $a_{\pm}=\frac{3c^{2}}{2\left|\Lambda\right|}\left(1\pm\sqrt{1+\frac{4k}{3c^{4}}\left| \Lambda \right|}\right)$.

\begin{figure}[th]
\includegraphics[scale=0.4]{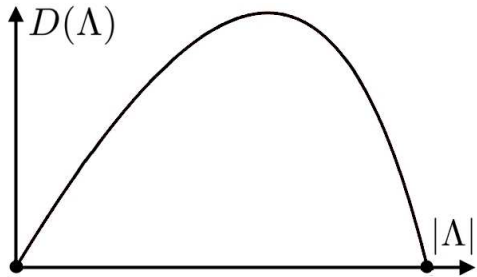}
\caption{The graph of the function $D(\Lambda)$, given by Eq.(\ref{Difference}), indicates that a positive spatial
curvature leads to two solutions: $\Lambda=0$ and $\Lambda<0$.
The latter corresponds to ``anti-Einstein" static and a closed Universe with a negative cosmological
constant given by Eq.(\ref{NegLam}).}
\label{fig3}
\end{figure}

\bigskip

(iii2) \textit{Is $\Lambda=0$ possible?}

\bigskip

In this case the corresponding equation of motion for the scale factor takes the form
\begin{equation}
           \dot{a}^{2}-c^{2}a=-k.
\label{Potential5}
\end{equation}

(iii2.1) $k>0$; \quad allowed

(iii2.2) $k=0$; \quad allowed

(iii2.3) $k<0$; \quad allowed

\bigskip

One can easily verify that this case presents no restriction on $k$
and that any solution of Eq.(\ref{Potential}) is consistent with the functional equation Eq.(\ref{GeneralEquationState}). 
We therefore conclude that in this case, the Universe is either spatially open, closed or flat.

\bigskip

\begin{center}
\textbf{Normalized action principle for a perfect fluid with a particle number current term}
\end{center}
  
  \bigskip

We now consider a somewhat different and more elaborate formulation of the perfect fluid normalized cosmology.
This approach followed mainly because it offers the possibility of considering the action principle of the
relativistic perfect fluid in the most general case, that is, not necessarily maximally symmetric.
In such a formalism, the proper pressure and the baryon number equation are explicitly introduced,
and one is not forced to import them indirectly through the corresponding energy-momentum conservation
equation Eq.(\ref{Conservation}).
Similar to our discussion above, we will derive the corresponding consistency
condition in the most general, not necessarily maximally symmetric, case. 
The maximal symmetry assumption is considered later as a particular limit case.
While reconfirming our previous main conclusions of a non-closed universe and a vanishing cosmological
constant, we also conclude that the kinematical condition given by Eq.(\ref{GeneralEquationState}) is now modified to 
Eq.(\ref{ConsistencyPFEqState}).
This should not come as a surprise to us, since it merely reflects the fact that the more general model described
by Eq.(\ref{NormalizedPF}) contains an additional current term in the action. Consequently, the value of this action
along the solution, which serves to fix the associated value of the cosmological constant,
receives an additional contribution and the corresponding functional equation is modified.
A few examples, not directly related to the problem discussed in this paper but
that demonstrate how additional terms in the normalized action may modify the value of
the cosmological constant, can be found in \cite{NGR}.

Let us consider the following normalized general relativistic action with a perfect fluid matter
\begin{equation}
\begin{split}
          &I_{C}\left[ g^{\mu \nu },j^{\mu },\theta ,\alpha ,\beta \right] = \\
          &\frac{\int d^{4}x\sqrt{-g}\left( -{\mathcal{R}}-j^{\mu }\left( \partial _{\mu
           }\theta +\alpha \partial _{\mu }\beta \right) -f\left( n\right) \right) }{\epsilon\int d^{4}x\sqrt{-g}},
\label{NormalizedPF}
\end{split}
\end{equation}
where the dynamical degrees of freedom are given by: $g^{\mu \nu }$, $j^{\mu }$, $\theta$, $\alpha$, $\beta$.
Here $n$ is the fluid particle number density that enters through the definition of the matter current 
\begin{equation}
            j^{\mu }=nu^{\mu }.
\label{DefCurrent}
\end{equation}
The latter is subject to the following normalization condition
\begin{equation}
           n^{2}=-j^{\sigma }j_{\sigma }  
\label{phi}.
\end{equation}
that follows from the normalization of the four-velocity vector $u^{\mu}u_{\mu}=-1$.
Performing the variation with respect to $\theta$ ensures that the particle number conservation equation 
\begin{equation}
          \nabla _{\mu }j^{\mu }=0
\end{equation}
holds.
 
Now let us consider the remaining equations of motion that stem from the action $I_{C}$.
Performing the variation with respect to $g^{\mu \nu }$ results in the equations
of motion, Eq.(\ref{NGReq}), where the yet unspecified cosmological constant $\Lambda$ is identified according to
\begin{equation}
           \Lambda=\frac{\epsilon}{2}I_{C}(\Lambda)
\label{EquationMetric}
\end{equation}
and the corresponding energy-momentum tensor $T_{\mu\nu}$ admits the structure
of a perfect fluid
\begin{equation}
           T_{\mu \nu }=pg_{\mu \nu }+\left( \rho +p\right) u_{\mu }u_{\nu }.
\label{PFEMT}
\end{equation}
The associated energy density $\rho$ and the pressure $p$ are given by 
\begin{equation}
\begin{array}{c}
          \rho =f\left( n\right)  
\\ 
           p=nf^{\prime }\left( n\right)-f\left( n\right),
\label{dictionary}
\end{array}
\end{equation}
and emerge after we have utilized the equations of motion
\begin{equation}
\begin{array}{c}
         \partial _{\mu }\theta -j_{\mu }\frac{f^{\prime }\left( n\right) }{n}=0 
\\ 
          u^{\mu }\partial _{\mu }\alpha =0 
\\ 
         u^{\mu }\partial _{\mu }\beta =0,
\end{array}
\label{EqMotion}
\end{equation}
that follow from varying $j^{\mu}$, $\beta$, $\alpha$, respectively.

We turn now to the derivation of the specfic form of the functional equation,
Eq.(\ref{EquationMetric}), for $\Lambda$ in the present case.
To this end, we trace the equation of motion Eq.(\ref{NGReq}) with $\Lambda$ given by Eq.(\ref{EquationMetric}) 
and contract the expression in Eq.(\ref{EqMotion}) 
with $j^{\mu }$ to obtain the following two equations
\begin{equation}
\begin{split}
           -{\mathcal{R}}&-2I_{C}-\frac{1}{2}\left( \rho -3p\right) =0
\\
            &j^{\mu }\partial _{\mu }\theta+\rho+p =0.
\end{split}
\label{JmuTheta}
\end{equation}
Plugging the corresponding ${\mathcal{R}}$ and $j^{\mu }\partial _{\mu}\theta$ 
back into the action Eq.(\ref{NormalizedPF}) we obtain
\begin{equation}
           I_{C}=\frac{\int d^{4}x\sqrt{-g}\left( 2\epsilon I_{C}+\frac{1}{2}\left( \rho -3p\right)
          -\rho+\overbrace{\rho+p}^{-j^{\mu }\partial _{\mu }\theta} \right) }{\epsilon \int d^{4}x\sqrt{-g}}
\label{PerfectFluidI}
\end{equation}
that acquires the form
\begin{equation}
          \Lambda=\frac{1}{4}\left\langle p(n)-\rho(n) \right\rangle,
\label{ConsistencyPF}
\end{equation}
after we have utilized Eq.(\ref{EquationMetric}) and the expression in Eq.(\ref{dictionary}).
Notice that omitting the contribution of $j^{\mu }\partial _{\mu }\theta$ to the action Eq.(\ref{PerfectFluidI}) would have resulted
in the consistency condition given by Eq.(\ref{LambdaRhoP}).
Furthermore, note that Eq.(\ref{ConsistencyPF}) has been derived for an arbitrary underlying spacetime symmetry
and without specifying a concrete perfect fluid equation of state.  
Assuming now a perfect fluid equation of state of the form in Eq.(\ref{PFEqState})
leads a differential equation for $f(n)$ that yields
\begin{equation}
          f\left( n\right) \propto n^{1+w}.
\label{f}
\end{equation}
In this case the functional equation, Eq.(\ref{ConsistencyPF}), acquires the form
\begin{equation}
          \Lambda=\frac{w-1}{4}\left\langle \rho(n) \right\rangle
\label{LambdaRhoN}
\end{equation}
where $\rho(n)$ is given by Eq.(\ref{f}).
Since the parameter $w$ is usually bounded between plus and minus unity we see from the latter and form Eq.(\ref{ConsistencyPFEqState})
that the corresponding solution for $\Lambda$, if exists, is non-positive.
In the following we consider the solution of the functional equation for $\Lambda$ in the maximally symmetric case. 

We now investigate the particular form the consistency condition Eq.(\ref{ConsistencyPF}) acquires, once
we consider the minisuperspace model, i.e., restrict our attention to metrics of the form Eq.(\ref{T}).
To this end, we have to plug the equations of motion into Eq.(\ref{ConsistencyPF}) and exclude the combination $p(a)-\rho(a)$.
In such case it is useful to consider the corresponding equations of motion, Eq.(\ref{NGReq}), when written in the form
\begin{equation}
          \mathcal{R}_{\mu\nu}=\frac{1}{2}\left(T_{\mu\nu}^{(full)}-\frac{1}{2}g_{\mu\nu}T^{(full)}\right)
\label{EinsteinDifferent}
\end{equation}
with
\begin{equation}
         T_{\mu\nu}^{(full)}=T_{\mu\nu}-2\Lambda g_{\mu\nu}.
\end{equation}
In fact, the $time-time$ and the $space-space$ components of Eq.(\ref{EinsteinDifferent}) are given by the following:
\begin{equation}
\begin{split}
          & -3\frac{\ddot{a}}{a}=\frac{1}{4}\left(\rho+3p-4\Lambda\right)
\\
          \frac{\ddot{a}}{a}+2&\left(\frac{\dot{a}}{a}\right)^{2}+2\frac{k}{a^{2}}=\frac{1}{4}\left(\rho-p+4\Lambda\right),
\end{split}
\label{SpaceSpaceEq}
\end{equation}
respectively, where the second expression in Eq.(\ref{SpaceSpaceEq}) is utilized to exclude the $p-\rho$ combination from Eq.(\ref{ConsistencyPF}) and
results in
\begin{equation}
           \left\langle\frac{\ddot{a}}{a}+2\left(\frac{\dot{a}}{a}\right)^{2}+2\frac{k}{a^{2}} \right\rangle=0.
\label{LambdaRhoRhoPrimeWithJ}
\end{equation}
Without the current term Eq.(\ref{LambdaRhoRhoPrimeWithJ}) differs from our previously derived condition Eq.(\ref{Kinematic}).
As expected, the condition Eq.(\ref{Kinematic}) may also be obtained once we substitute the first expression in Eq.(\ref{SpaceSpaceEq})
into the action given by Eq.(\ref{PerfectFluidI}) without the source term.

Similar to our discussion above in the minisupersymmetric case, we focus on a perfect fluid equation of state of the type in Eq.(\ref{PFEqState}),
and plug the corresponding energy density Eq.(\ref{rhoPF}) into Eq.(\ref{LambdaRhoN}), which yields the functional equation 
for $\Lambda$:
\begin{equation}
          \Lambda=\frac{3c^{2}(w-1)}{2}\left\langle a(t;\Lambda)^{-3(1+w)} \right\rangle.
\label{ConsistencyPFEqState}
\end{equation}
Keeping in mind that the parameter $w$ is usually bounded between plus and minus unity, we should seek 
non-positive solutions of Eq.(\ref{ConsistencyPFEqState})
for $\Lambda$ by focusing on the $\Lambda<0$ and the $\Lambda=0$ possibilities. 
Following the same steps as above, it is possible to show that Eq.(\ref{ConsistencyPFEqState}) 
admits solutions only in the cases of the generally vanishing cosmological constant and the non-closed Universe. 
Below we present our final results and a short discussion.

\bigskip

\textit{Is $\Lambda=0$ possible?}

\bigskip

It can be straightforwardly verified that the conclusions we derived in cases (i1), (ii) and (iii2), for different
equations of state hold in the present case as well. Specifically, the modified functional condition given by Eq.(\ref{ConsistencyPFEqState}) 
still respects the $\Lambda=0$ solution with $k \leq 0$, and forbids the $k>0$ case, once the
corresponding solution of the equation of motion Eq.(\ref{Motion}) is plugged in.
One exception of this general behavior is the $w=1$ choice, which corresponds to the so-called stiff matter case with $p=\rho$.
In this case there is no restriction on the sign of $k$ and the spatial curvature of the spacetime.

\bigskip

\textit{Is $\Lambda<0$ possible?}

\bigskip

Considering graphs of various $D(\Lambda)$, similar to Eq.(\ref{Difference}), for the equations of state that 
correspond to the cases (i), (ii) and (iii), one may show that the negative cosmological constant 
$\Lambda<0$ is not compatible with Eq.(\ref{ConsistencyPFEqState}).

To summarize, we have shown that starting from FRW cosmology
in the framework of normalized general relativistic theory with matter subject to the strong energy
condition led us generically to a non-closed Universe. The corresponding
cosmological constant associated with this type of matter is generically
zero, except in non-generic solutions associated with Einstein
static closed and Eddington-Lema\^{\i}tre universes \cite{Eddington}.
Following the perfect fluid action principle \cite{Jackiw}, we learn that the inclusion 
of the perfect fluid particle current term in the action excludes Eddington-Lema\^{\i}tre universes 
and leads to a vanishing cosmological constant and a non-closed Universe. 

\bigskip
      \textbf{Acknowlegdments}
\bigskip

The authors cordially thank Ilya Gurwich and Eduardo Guendelman for their useful
comments.

\end{document}